%% file: _main.tex
\documentclass[letterpaper, 10 pt, conference]{ieeeconf}
\IEEEoverridecommandlockouts                              
\usepackage{microtype}
\usepackage{graphicx}
\usepackage{subcaption}
\usepackage{booktabs} 

\usepackage{hyperref}
\usepackage{algorithm}
\usepackage{algorithmic}
\usepackage{amsmath}
\usepackage{amssymb}
\usepackage{mathtools}

\usepackage[disable,textsize=tiny]{todonotes}
\presetkeys{todonotes}{inline}{}

\providecommand{\citep}[1]{(\cite{#1})}

\title{\large \bf Model-based adaptation for sample efficient transfer in reinforcement learning control of parameter-varying systems}

\author{Ibrahim Ahmed$^{1,4}$ and Marcos Quinones-Grueiro$^{2,4}$ and Gautam Biswas$^{3,4}$
    \thanks{*This work was supported by NASA Award \#80NSSC21M0087-21-S06.}
    \thanks{$^{1}${\tt\small ibrahim.ahmed@vanderbilt.edu}}%
    \thanks{$^{2}${\tt\small marcos.quinones.grueiro@vanderbilt.edu}}%
    \thanks{$^{3}${\tt\small gautam.biswas@vanderbilt.edu}}%
    \thanks{$^{4}$Institute for Software Integrated Systems, Vanderbilt University, Nashville, TN, USA}%
}

\begin{document}
\maketitle
\thispagestyle{empty}
\pagestyle{empty}

\begin{abstract}
In this paper, we leverage ideas from model-based control to address the sample efficiency problem of reinforcement learning (RL) algorithms. Accelerating learning is an active field of RL highly relevant in the context of time-varying systems. Traditional transfer learning methods propose to use prior knowledge of the system behavior to devise a gradual or immediate data-driven transformation of the control policy obtained through RL. Such transformation is usually computed by estimating the performance of previous control policies based on measurements recently collected from the system. However, such retrospective measures have debatable utility with no guarantees of positive transfer in most cases. Instead, we propose a model-based transformation, such that when actions from a control policy are applied to the target system, a positive transfer is achieved. The transformation can be used as an initialization for the reinforcement learning process to converge to a new optimum. We validate the performance of our approach through four benchmark examples. We demonstrate that our approach is more sample-efficient than fine-tuning with reinforcement learning alone and achieves comparable performance to linear-quadratic-regulators and model-predictive control when an accurate linear model is known in the three cases. If an accurate  model is not known, we empirically show that the proposed approach still guarantees positive transfer with jump-start improvement.
\end{abstract}

\input{sections/introduction}
\input{sections/problem}
\input{sections/relatedwork}
\input{sections/proposal}
\input{sections/experiments}
\input{sections/conclusions}

\bibliographystyle{IEEEtran}
\bibliography{ref/references,ref/supp}

\end{document}

%% file: sections/introduction.tex
\section{Introduction}
\label{sec:intro}

Transfer learning in control seeks to reuse knowledge gained from past source tasks to speed up or improve performance on related, target tasks. This is an especially attractive proposition in applications where time or training samples are sparse. For example, consider data-driven fault-tolerant control, as surveyed by \cite{li2022perspective}, where a controller needs to adapt quickly and sufficiently well to a changed environment.

However, the questions underlying transfer are: \textit{what knowledge to transfer, and how do we transfer it well?} Poorly related tasks may cause negative transfer when conventional transfer learning methods are applied, especially if the relationship between tasks is not considered in the transfer process \cite{wang2019characterizing,cao2010adaptive}.
Fundamentally, the performance of transfer learning between tasks is dependent on the relationships between tasks. Tasks that are similar will have similar control policies, therefore will need lesser time and data for their policies to adapt to each other.

Thus far, a substantial body of work has addressed transfer in the data and algorithm space. Task similarity measures have been developed from the statistical properties of measurements across tasks. Stochastic algorithms have been proposed (sections \ref{sec:similarity_transfer}, \ref{sec:learning_optimization}) that leverage model architectures, machine learning hyperparameters, and control policy parameter update rules to achieve faster, jump-start, or asymptotically higher performance improvement on the target task, all the while being agnostic to the underlying process dynamics.

This work makes contributions in a related direction. We address transfer in the space of process dynamics. 
While our approach is applicable to a broad class of dynamical systems, we demonstrate strong theoretical results for systems with linear, time-invariant dynamics. By modeling the relationships between source and target task dynamics as linear transformations, we develop a transformation of the source policy to transfer on the target task. We demonstrate conditions where the transformation will produce behavior  optimal in the least squares sense. Generally, for approximately identified target tasks or locally optimal source policies, the transformation may be used as a policy initialization to get a jump-start improvement, prior to further optimization.

The following section describes the transfer learning problem for this work in the context of reinforcement learning. Following that, in section \ref{sec:related_work}, we review different approaches toward transfer. Section \ref{sec:approach} motivates cases for invariance of optimal policies and cases for the derivation of policy transforms. Finally, experiments are done on dynamical systems using stochastic and classical control approaches to demonstrate these concepts.

%% file: sections/problem.tex
\section{The transfer learning control problem}
\label{sec:problem}

In this work, we discuss the transfer of control across systems modeled as Markov Decision Processes (MDPs). An MDP control problem is a task $T \in \mathcal{T}$, characterized by the process dynamics $P: X \times U \rightarrow X$, which map the current state $x_i$ to the next state $x_{i+1}$, given an input $u_{i+1}$. Each state transition is assigned a reward based on the state and the action taken to reach there $r: X \times U \rightarrow \mathbb{R}$. An episode is a sequence of interactions until a terminal or goal state is reached. The objective for $T$ is to derive a policy $\pi_T: X \rightarrow U$, such that each action picked maximizes expected future returns $\mathbb{E}[G(x_i)]$, where $G(x_i) = \sum_{j=i}\gamma^{j-i}r(x_i,\pi_T (x_i))$ is the discounted sum of rewards starting at $x_i$ following some policy. The discount factor $\gamma \in (0,1]$ prioritizes the immediacy of feedback. Expected returns under an optimal policy are known as its value, $V(x)=\max_{u}\mathbb{E}[G(x)]$. Then, $\pi_t(x) \gets \arg \max_u V(x)$.

The transfer problem is summarized as follows. A task $T$ consists of the process dynamics $P$ and the reward function $r$. Given a target task $T_t$ and a population of source tasks $\mathcal{T}_s \in \mathcal{T}$, find a source task $T_s \in \mathcal{T}_s$ and a transfer mechanism, such that the performance of its policy fine-tuned on $T_t$ provides an optimal solution for $T_t$. In other words,

\begin{align}
T_s: \max_{T \in \mathcal{T}_s} G(x \sim T_t \mid \pi_{T\rightarrow T_t})
\label{eqn:objective}
\end{align}

$T_s$ and $T_t$ may differ on process dynamics and reward. For example, due to a fault or a changed control objective. Both affect evaluated returns $G$. The objective in equation \ref{eqn:objective} can only be achieved retroactively, when candidate source tasks have been evaluated on the target task. However, exploring with transfer of each source task may violate time and safety constraints in specific applications. Therefore, the challenge is to \textit{preemptively} select a favorable source task and transfer mechanism. For this work, we assume homogeneous transfer: that the state and action spaces of all tasks are identical. Practically, this applies to cases where a single MDP's state transitions are disturbed due to faults, degradation etc. For the remainder of this work, subscripts $\ast_s, \ast_t$ refer to source and target parameters respectively.

The goal, ultimately, is to achieve high returns on the target task. This may be done by either picking a source task liable to transfer well, or by tweaking the transfer process such that the source policy converges swiftly to an optimum on the target task. Prior related work has addressed both of these approaches. 

%% file: sections/relatedwork.tex
\section{Related Work}
\label{sec:related_work}

This section reviews prior work done in this and related fields. The aforementioned goal of transfer learning for control overlaps with multi-task learning \cite{zhang2021survey}, reward shaping \cite{brys2015policy}, and few-shot learning \cite{wang2020generalizing}. The body of research is divided into two broad categories: optimizing transfer learning informed by task similarity, and optimizing transfer once a source task is picked.

\subsection{Selecting Similar Tasks for Transfer}
\label{sec:similarity_transfer}

A multitude of similarity measures have been used in transfer learning looking at task similarity in classification and regression problems. Work by \cite{zhou2021task} builds on top Model-Agnostic Meta-Learning (MAML) \cite{finn2017model} to develop task similarity-aware MAML. They represent task similarity as Euclidean distance between parameters of models trained on sampled tasks. Clusters are made for similar tasks, and the cluster closest to the target task is used during meta-initialization.
\cite{fernandes2019hypothesis} propose a general transfer approach, where task similarity is used to regularize the transferred model.
\cite{wang2019transfer} formalize the intuition that similar tasks have similar performances, and use the performance gap as a regularizer for transfer learning. Euclidean distance between model coefficients is used for this. 
Alternatively, \cite{zhang2017effect} hypothesize that, given samples, two (classification) tasks are similar if the accuracy on the source task and the target tasks is high.
\cite{lu2017discriminative} propose a reconstruction classifier, which attempts to reconstruct samples from the target task using samples from the source task. The assumption being that samples from $T_t \cup \mathcal{T}_s$ lie on a subspace and can be modeled as combinations of each other. The sparser the coefficients for reconstruction, and the lower the error, the higher the similarity.
Whereas \cite{shui2019principled}  looks at the utility of explicitly including task similarity measures in the learning process. They use $\mathcal{H}$ divergence and Wasserstein distance between probability distributions in adversarial multi-task learning for classification tasks.
Recently, \cite{alam22} develop physics-guided models to develop an initial reinforcement-learned control policy under inaccurate rewards, which in turn is transferred to the target process. The initialization of the source policy reduces the sample size needed to adapt to the target task.

\subsection{Optimization of the Learning Process}
\label{sec:learning_optimization}

Another class of approaches has addressed the learning process itself, once a source task is selected. Such methods are not restricted to transfer learning, but are applicable to learning algorithms in general. Relevant work in this area touches on meta-learning \cite{hospedales2021meta}, neural architecture design \cite{du2019transfer,luo2019hierarchical}, and hyperparameter tuning \cite{yogatama2014efficient,paul2019fast}.

In recent work in classical control, \cite{chakraborty2020theoretical} propose the design of a transferrable controller via system identification. They stimulate the target process with tailored exploratory actions from the same initial state to identify relationships with the source process. The source policy can then be transformed to the identified target process.

%% file: sections/proposal.tex
\section{Policy Transfer via Transformation}
\label{sec:approach}

In this section, we derive a policy transformation for a category of tasks such that the source policy is as close to optimal in the target task in the least squares sense, assuming the same reward function.

The optimal policy $\pi$ - the control law - is derived to reach the optimal set of states \textit{and} using optimal actions, since the reward is a function of both state and action. However, if the reward is only a function of performance ($x$), the optimal policy would depend only on the states traversed, not the effort ($u$) it took to traverse them.

\begin{align}
    &r: X \rightarrow \mathbb{R} \implies \nabla_{u}r=0 \nonumber \\
    &\pi: \arg \max_{\pi} \mathbb{E}_{x_i \sim X}\sum_{j=i}^\infty \gamma^{j-i}r \left(P(x_i, \pi(x_i)) \right) \nonumber \\
    &\nabla_P V(x)=0\label{eqn:only_state}
\end{align}

That is, the optimal states remain the same. The same set of states optimal under $T_s$ are optimal under $T_t$. This comes with one caveat: $P_t$ is controllable for the same set of states \citep{sussmann1972controllability}. Meaning, the dynamics in $P_t$ allow those optimal states to be reached by some actions. Therefore if $\pi_t$ can bring about the same state change in $P_t$ as $\pi_s$ does in $P_s$, $\pi_t$ can be guaranteed to be optimal.

Reward being a function of state only is a strong assumption. However, it is possible in many cases. For instances, where the cost of actions is dwarfed by the incentive to drive the state variables to a certain point. In real world scenarios such as vehicle navigation, the battery or fuel levels may be included in the state vector as a proxy for action magnitudes.


To that end, sequential MDPs are framed as deterministic, dynamical time-invariant processes from a control theory lens. A process $P$ has a state $x \in \mathbb{R}^n$, and inputs $u \in \mathbb{R}^m$. The process itself is characterized by the rate of change of its state $\dot{x} = F_{\dot{x}}: X \times U \rightarrow X$. For systems with linear dynamics, $F_{\dot{x}}=Ax + Bu$, where $A \in \mathbb{R}^{n\times n},B \in \mathbb{R}^{n\times m}$ are constant matrices. $A$ is the response to internal state, and $B$ is the response to external inputs.

\begin{align}
    \dot{x}_i &=F_{\dot{x}} (x_i, u_{i+1}) \nonumber \\
    x_{i+1} &= P(x_i, u_{i+1} \mid F_{\dot{x}}) \nonumber \\
            &= x_i + \int_{t=i}^{t=i+1}\dot{x}_i\partial t \nonumber \\
            &\approx x_i + \delta t \cdot \dot{x}_i
    \intertext{Where $\delta t$ is a discrete sampling interval. For linear systems, $P$ can be approximated as a linear transformation,}
    \dot{x}_i &= A x_i + B u_{i+1} \nonumber \\
    x_{i+1}&\approx P \cdot [x_i, u_{i+1}]^T \nonumber \\
            &=[1 + \delta t A, \delta t B ][x_i, u_{i+1}]^T \label{eqn:dynamics}
\end{align}

We assume $F_A: \mathbb{R}^{n\times n}\rightarrow \mathbb{R}^{n\times n}, F_B: \mathbb{R}^{n\times m} \rightarrow \mathbb{R}^{n\times m}$ are transformations of $A_s, B_s$ under a fault, and $\pi_s$ is the control policy of the source task. The control policy $\pi_s$ already has optimized inputs to change states to optimal positions. Since the optimal states remain invariant, we want a policy $\pi_t$ such that the change of state under the target process is the same as the source process.

\begin{align}
    &\dot{x}_s = A_s x + B_s u_s  \nonumber \\
    &\dot{x}_t = F_A A_s x + F_B B_s u_t  \nonumber  \\
    & \text{If }\dot{x}_t = \dot{x}_s  \text{, then,}\nonumber  \\
    &\implies u_t = (F_B B_s)^{-1}(I - F_A)A_s x + (F_B B_s)^{-1}B_s u_s  \label{eqn:policy_xform_add}\\
    &\implies u_t =\left((F_B B_s)^{-1}(I-F_A)A_s  + (F_B B_s)^{-1}B_s \pi_s\right)x  \label{eqn:policy_xform_mul}
\end{align}

\begin{figure}[ht]
\centering
\includegraphics[width=0.3\textwidth]{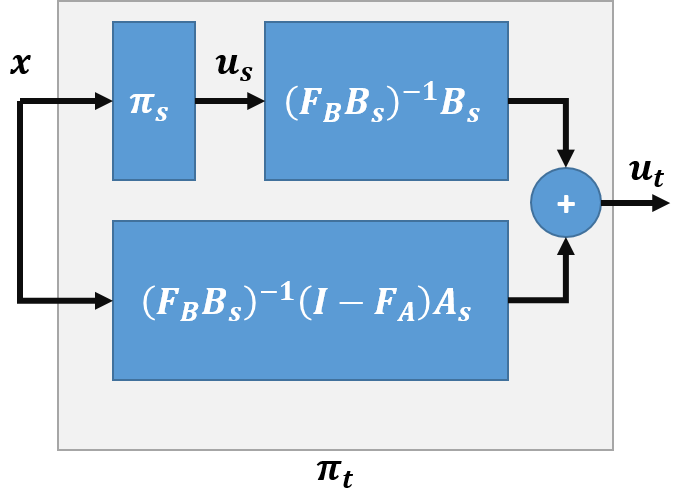}
\caption{A schematic of the policy transformation. The nominal action $u_s$ is transformed as $u_t$ to bring about the same change in state in the target system that was considered optimal by $\pi_s$ in the source system. The source policy can be any control algorithm for e.g. LQR, MPC, RL etc.}
\label{fig:schematic}
\end{figure}

Equation \ref{eqn:policy_xform_add} represents a multiplicative transformation of the source policy by $(F_B B_s)^{-1}B_s $, representing a change in the input's effect on state dynamics. And an additive correction by $(F_B B_s)^{-1}(I - F_A)A_s$, representing the changed internal dynamics of the system. Equation \ref{eqn:policy_xform_mul} factors out $u_s \gets \pi_s x$ to get an equivalent representation for $u_t$. Figure \ref{fig:schematic} depicts how these transformations can be appended in series and parallel, respectively to an existing policy function $\pi_s$.

The transformation will change the policy's response to each state. The range of new actions may be different from that of the source policy. If the target policy's range is a subset of the source's range, or if actions are unconstrained in the process, this will not be notable. For constrained actions, actions can be clipped. For linear systems, this will not affect their eventual dynamics. That is, a sequence of actions scaled down by a factor of $k \in \mathbb{R}^+$ over $k$ intervals will result in the same change in state as the unscaled action applied for one interval.

The transformation of state dynamics, $F_A$ is not required to be invertible. The only functions needing inversion are the action dynamics $B_s$ and their transformation $F_B$. In cases where a perfect inverse does not exist, the Moore-Penrose Pseudo-inverse may be used as the closest approximation in the least squares sense. The approximation error of the pseudo-inverse may be a measure of the suitability of the transformed policy.

$A_s,B_s,F_A,F_B$ are learned through system identification strategies. For the specific case where process dynamics $P(x,u \mid A,B)$ and their transformations $F_A, F_B$ can be linearized in the region of interest, they can be learned from measured data by solving a least squares problem:

\begin{align}
    X &= [x_0, x_1, ...], X^+ = [x_1, x_2, ...], U = [u_1, u_2, ...] \nonumber \\
    P &= \left[X^+ \cdot [X; U]^T\right] \cdot \left([X; U] \cdot [X; U]^T\right)^{-1} \label{eqn:dynamics-solve}
\end{align}

Then, $P$ can be decomposed and solved for $A, B$ as in equation \ref{eqn:dynamics}. If $A_s, A_t$ are known, then $F_A = A_t \cdot A_s^{-1}$. Similarly, for $F_B, B_s, B_t$. Since $B$ is not necessarily square, $F_B = (B_t \cdot B_s^T) \cdot (B_s \cdot B_s^T)^{-1}$. The whole approach is outlined in algorithm \ref{alg:algorithm}.

\begin{algorithm}[tb]
\caption{Policy transformation via system identification.}
\label{alg:algorithm}
\textbf{Input}: $\pi_s, T_s, T_t$\\
\textbf{Parameter}: Data buffer $\mathcal{D}_s, \mathcal{D}_t$\\
\textbf{Output}: $\pi_t$
\begin{algorithmic}[1] 
\STATE Collect $(x_i, u_{i+1}, x_{i+1})$ into $\mathcal{D}_s$ using $P_s, \pi_s$
\STATE Collect $(x_i, u_{i+1}, x_{i+1})$ into $\mathcal{D}_t$ using $P_t, \pi_s$
\STATE Use equations \ref{eqn:dynamics-solve}, \ref{eqn:dynamics} to learn $A_s, B_s, F_A, F_B$
\STATE Use equation \ref{eqn:policy_xform_mul} to get $\pi_t$
\STATE \textbf{return} $\pi_t$
\end{algorithmic}
\end{algorithm}

For non-linear transformations of the linear system matrices $A,B$, non-linear basis functions such as neural networks can be used to approximate the target transformation. Since $F_B B_s$ requires inversion, a monotonic constraint should be put on their learned models (by constraining hidden layer weights to be positive, for example). Or, probabilistic models which learn a posterior distribution of a variable, given the output of a function \cite{ardizzoneanalyzing} can be used, from which the likely inverse of the function can be sampled.

In summary, if reward is a function of only the state, and if optimal states in $P_s$ are reachable in $P_t$, then optimal change in state remains invariant. Thus $\pi_t$ can be derived as a transformation of $\pi_s$ in terms of $P_s$ and $P_t$ to bring about the same change in state to optimize the target task.

The worst-case computational cost of our approach is lower than that of approaches that will evaluate states and actions \textit{anew} on the target task, such as RL and MPC. Our approach relies on system identification to transform $\pi_s$, which is already known. As an illustration, we consider deterministic, continuous MDPs, assuming unique actions lead to unique states. Identifying an arbitrary system $P$ requires sampling each state transition once to learn that mapping $x_{i+1} \gets P(x_i, u_{i+1})$. The complexity of identification then is the space $X \in \mathbb{R}^n \times U \in \mathbb{R}^m$. The Bellman equation \citep{bellman1966dynamic}, which is foundational to RL and MPC approaches, traverses all possible state \textit{trajectories} to evaluate states, where state transitions may be traversed multiple times, thus giving a higher computational complexity. Therefore, learning the target dynamics to transform the source policy is computationally cheaper than evaluating states in the target task. Even when a source RL policy is re-used and fine-tuned on the target task, it may yet need to sample every state trajectory in the worst case. The source policy has learned to drive towards valuable states under $P_s$. It needs to re-sample the transformed trajectories under $P_t$ to reevaluate each state. The amount of trajectories needing revision will measure, and depend on, the similarity between $P_s$ and $P_t$.

%% file: sections/experiments.tex
\section{Experiments}

This approach is demonstrated using linear and non-linear systems. We demonstrate our case with Linear Quadratic Regulator (LQR) \citep{lavretsky2013optimal}, Model-Predictive Control (MPC), and RL. For RL, the Proximal Policy Optimization (PPO) algorithm is applied. The reward $r$ and cost $c$, where $r=-c$, are specified as quadratic functions of state $x$ with weights $Q$, where $Q$ is a diagonal matrix. The function minimizes cost and maximizes reward around $x=\hat{0}$, which is the desired state vector. However, the system can be driven to some other point $x_0$ by substituting $x \gets x-x_0$ without a loss in generality. For the sake of comparison with LQR, we introduce a small action weight $R=10^{-5}$ in reward, which otherwise does not affect other approaches. Similarly, to accommodate LQR, the optimization assumes unconstrained actions. However, during testing, the actions are clipped to $[-1, 1]$ for each time interval.

\begin{align}
    c = -r = x^T Q x + u^T R u
\end{align}

A simple one-dimensional temperature ($x$) regulation system is first used as a test bed, where positive and negative actions control a heating or cooling element ($u$). System parameters are set as $a=-0.1, b=1, Q=I$. Faults represent a change in conductivity $a$, and a reversal in action polarity $b$, such that the nominal action of increasing heat will now cool the system.

\begin{align}
    &x,u,a,b, F_A, F_B \in \mathbb{R} \nonumber \\
    &\dot{x}_{temp} = ax + bu \label{eqn:sys_temp}
\end{align}

A higher dimensional, but still linear, spring-mass system is described in equation \ref{eqn:sys_spring}, with $Q=I$, and actions $u$ as forces. The dynamics are governed by the mass $m=1$, spring constant $k=10$, and dynamic friction $k_f=0.2$.

\begin{align}
&x \in \mathbb{R}^2,\; u \in \mathbb{R},\; k,k_f,m\in \mathbb{R}^+, F_A, F_B \in \mathbb{R}^{2\times2} \nonumber \\
&\dot{x}_{spring}=
\begin{bmatrix}0 & 1 \\ -\frac{k}{m} & \frac{k_f}{m} \end{bmatrix}
x + 
\begin{bmatrix}0 \\ \frac{1}{m}\end{bmatrix}
u \label{eqn:sys_spring}
\end{align}

A yet more complex non-linear, continuous system is a pendulum in equation \ref{eqn:sys_pendulum}, with $Q=I$, and actions $u$ as torques. The dynamics are governed by the mass $m=0.1$, pendulum length $l=1$, gravitational acceleration $g=10$, and dynamic friction $k_f=0.02$.

\begin{align}
&x \in \mathbb{R}^2,\; u\in \mathbb{R},\; g,l,k_f,m\in \mathbb{R}^+, F_A, F_B \in \mathbb{R}^{2\times2} \nonumber \\
&\dot{x}_{pendulum} = 
\begin{bmatrix}0 & 1 \\ -\frac{g\sin(\cdot)}{l} & -\frac{k_f}{ml^2} \end{bmatrix}
x + 
\begin{bmatrix}0 \\ \frac{1}{ml^2}\end{bmatrix}
u \label{eqn:sys_pendulum}
\end{align}

Finally, a cartpole system is used to demonstrate a more complex non-linear case. The state vector $x$ comprises of the angle of the pole $x_1$, its angular velocity $x_2$, the position of the cart $x_3$, and the cart's velocity$x_4$. Actions $u$ are forces applied to the cart. The system is parametrized by the cart and pole masses, $m_c=0.5, m_p=0.1$, the length of the pole $l=1$, gravitational acceleration $g=10$, and the coefficient of friction $k_f=0.01$. The state weights for reward are $Q=[[1,0,0,0], [0,0.1,0,0],[0,0,10^{-5},0],[0,0,0,0.1]], R=10^{-5}$. For the reinforcement learning controller, each time step the pole is upright gains a constant reward of $1$. The state equations are factored into terms that are functions of $x$ and $u$, and $F_A, F_B \in \mathbb{R}^{4\times4}$ are applied as disturbances.

\begin{align}
\dot{x_1} &= x_2 \nonumber \\
\dot{x_2} & = \frac{g}{l}\sin x_1 - \frac{k_f x_2}{ml^2} + \frac{\dot{x_4}\cos x_1}{l} \nonumber \\
\dot{x_3} & = x_4 \nonumber \\
\dot{x_4} & = \frac{m_p \sin x_1\left(g\cos x_1-l\dot{x_1}^2\right) + u(t)}{m_c+m_p-m_p\cos^2 x_1}
\end{align}

Experiments are carried out by obtaining a source policy $\pi_s$ on the nominal process $P_s$. Then, a fault, denoted by a parametric change in the equations of state, is introduced. The change is characterized by $F_A$, a positive definite matrix, and $F_B$, a negative definite matrix. A buffer of measurements $\mathcal{D}_t$ is collected to estimate the target process $P_t$. The transformed policy $\pi_t$ is derived from $\pi_s$ by approximating both $P_s$ and $P_t$ as linear systems about the buffer.

We evaluate \textit{jumpstart improvement}, \textit{asymptotic improvement}, and \textit{time to threshold}. These metrics describe short and long term advantages of our approach and its computational complexity. Jumpstart improvement is the immediate difference in rewards when a policy interacts with a new task. Asymptotic improvement is the limit of accumulated rewards as the policy continues to learn. And time to threshold is the time taken for accumulated rewards to reach an acceptable level of performance.

Two sets of experiments are carried out. First, LQR and MPC are used with our approach.  They are model-based, deterministic, and have complete knowledge of task dynamics and the relationship to the target task. Thus, variables such as hyperparameter selection, system identification, and policy function formulation in RL are removed to make results of this work more apparent. $F_A, F_B$ are provided instead of estimated. The results are tabulated in table \ref{tbl:exp_tables_classical}. As both tables \ref{tbl:exp_lqr} and \ref{tbl:exp_mpc} show, the episodic rewards from $\pi_t$ (our transformation) are within a standard deviation of, if not better than, the benchmark $\pi_{lqr}$ and $\pi_{mpc}$.

In the second set of experiments, RL is applied (table \ref{tbl:exp_tables_rl}). For RL, $\pi_s$ is obtained by running PPO algorithm until the episodic rewards converge. The policy transformation applied to $\pi_s$ can be further fine-tuned via gradient descent. We represent the transformed policy as $\pi_t$, the source policy fine-tuned on $P_t$ as $\pi_s^*$, the policy fine-tuned with transformation parameters as $\pi_t^+$, and the one excluding parameters as $\pi_t^-$. In the first subset of experiments, the  $F_A, F_B$ are known \textit{a priori} (table \ref{tbl:exp_rl_known}). Then, the same experiments are repeated but where the transformations $F_A, F_B$ have to be learned from measured data (table \ref{tbl:exp_rl_unknown}). For both cases, $\mathcal{D}_s, \mathcal{D}_t$ use at most five episodes, amounting no more than $2,500$ interactions with the system. This will notably contrast with the time steps taken by RL to converge to a policy.

\begin{figure*}
     \centering
     \begin{subfigure}[b]{0.24\textwidth}
         \centering
         \includegraphics[width=\textwidth]{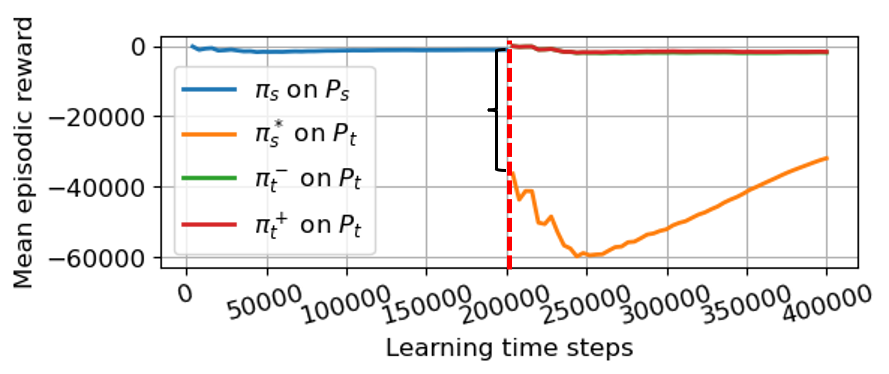}
         \caption{Temperature system.}
         \label{fig:rl_data_learn_temp}
     \end{subfigure}
     \hfill
     \begin{subfigure}[b]{0.24\textwidth}
         \centering
         \includegraphics[width=\textwidth]{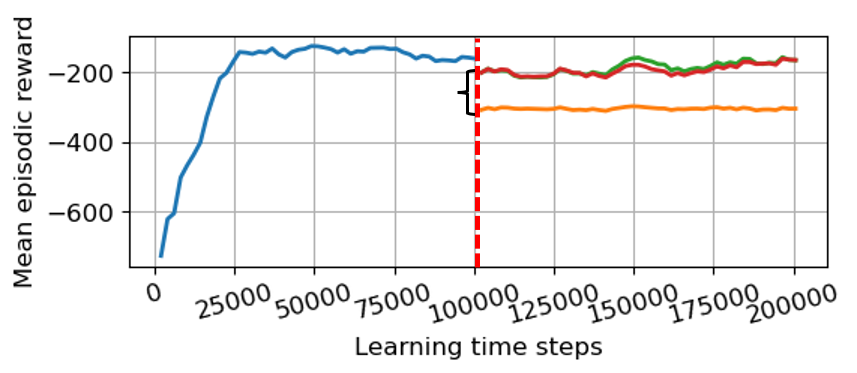}
         \caption{Spring system.}
         \label{fig:rl_data_learn_spring}
     \end{subfigure}
     \hfill
     \begin{subfigure}[b]{0.24\textwidth}
         \centering
         \includegraphics[width=\textwidth]{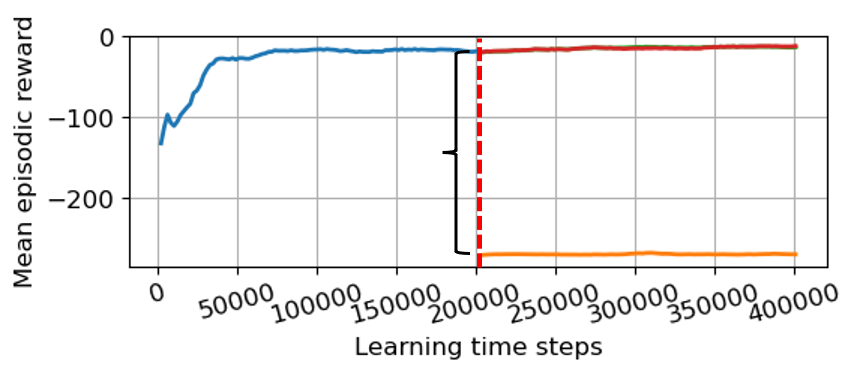}
         \caption{Pendulum system.}
         \label{fig:rl_data_learn_pendulum}
     \end{subfigure}
     \hfill
     \begin{subfigure}[b]{0.24\textwidth}
         \centering
         \includegraphics[width=\textwidth]{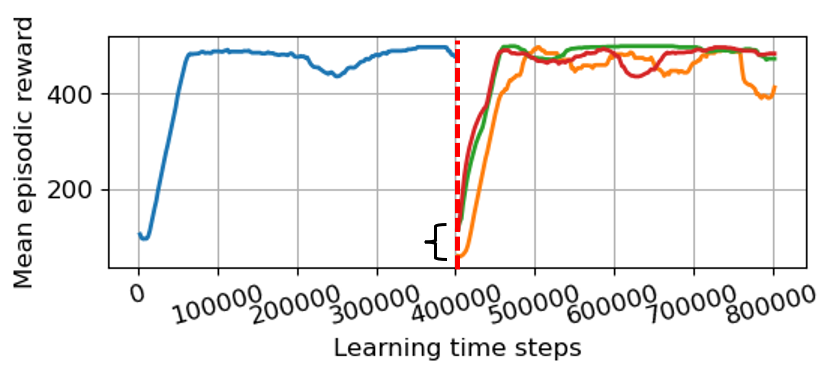}
         \caption{Cartpole system.}
         \label{fig:rl_data_learn_cartpole}
     \end{subfigure}
     \caption{Performance of RL trained on the source task $\pi_s$ (blue), transformed for the target task $\pi_t$, and later fine-tuned on target task with (red) and without (green) fine-tuning the transformation too, $\pi_t^+$/$\pi_t^-$. Performance is compared against continuing to learn $\pi_s$ to learn the target task (orange). The vertical line represents a fault occurrence.}
     \label{fig:learning}
\end{figure*}

Figure \ref{fig:learning} shows the effect the fault has on accumulated rewards, and how our policy transformation causes a jumpstart improvement as the RL policy tunes control after the system has changed. A parametric fault is introduced once RL has converged to a policy on $P_s$. There is an abrupt fall in rewards on $P_t$. Using RL iteratively to learn on $P_t$ is slow, and sometimes unable to recover at all. However, the policy transformation leads to a jumpstart improvement in rewards. For simple linear systems such as temperature, the transformation is instantly optimal. For non-linear systems like cartpole, there is a smaller jumpstart improvement, along with a faster time to convergence. For both sets of experiments involving RL, the results show that the transformed source policy, derived from the identified target system, is able to achieve comparable, if not better, performance than the source policy fine-tuned directly on the target task.

\begin{table*}
    \centering
    \footnotesize
    \begin{subfigure}{\textwidth}
        \centering
        \begin{tabular}{lllll} 
        \hline
                                & Temperature                   & Spring                & Pendulum              & Cartpole \\ 
        \hline
        $\pi_s$ on $P_t$        & $-51123.05\pm 17246$          & $-233.23\pm 53$       & $-271.66\pm 2$        & $40.4 \pm 3$\\ 
        \hline
        $\pi_{lqr}$ on $P_t$    & $\mathbf{-904.00 \pm 1624}$   & $\mathbf{-2.33\pm 2}$ & $\mathbf{-7.15\pm 8}$ & $\mathbf{500 \pm 0}$\\ 
        \hline
        $\pi_t$ on $P_t$        & $\mathbf{-904.01 \pm 1624}$   & $-4.22\pm 5$          & $-7.61\pm 8.2$        & $\mathbf{500 \pm 0}$\\
        \hline
        \end{tabular}
        \caption{Applying transformation to LQR. $\pi_t$ is compared with the policy $\pi_{lqr}$, derived directly on $P_t$. LQR policy is derived on the system linearized about equilibrium state.}
        \label{tbl:exp_lqr}
    \end{subfigure}
    \begin{subfigure}{\textwidth}
        \centering
        \begin{tabular}{lllll} 
        \hline
                                & Temperature               & Spring                & Pendulum              & Cartpole \\ 
        \hline
        $\pi_s$ on $P_t$        & $-36478.06\pm 15043$      & $-208.92\pm 67$       & $-23.53\pm 24$        & $79\pm 16$ \\ 
        \hline
        $\pi_{mpc}$ on $P_t$    & $-1138.50\pm 2075$        & $\mathbf{-2.82\pm 3}$ & $\mathbf{-7.29\pm 8}$ & $155 \pm 45$\\ 
        \hline
        $\pi_t$ on $P_t$        & $\mathbf{-533.23\pm 954}$ & $-4.56\pm 6$          & $-7.60\pm 8$          & $\mathbf{168 \pm 48}$\\
        \hline
        \end{tabular}
        \caption{Applying transformation to MPC. $\pi_t$ is compared with the policy $\pi_{mpc}$, derived directly on $P_t$ with a receding horizon of 5 time steps.}
        \label{tbl:exp_mpc}
    \end{subfigure}
    \caption{Mean episodic rewards using LQR and MPC, compared with $\pi_t$.}
    \label{tbl:exp_tables_classical}
\end{table*}

\begin{table*}
    \centering
    \footnotesize
    \begin{subfigure}{\textwidth}
        \centering
        \begin{tabular}{lllll} 
        \hline
                                & Temperature                   & Spring                    & Pendulum                  & Cartpole \\ 
        \hline
        $\pi_s$ on $P_t$        & $-51123.05 \pm 17246$         & $-295.39 \pm 70$          & $-271.66 \pm 2$           & $56.8 \pm 15.16$\\ 
        \hline
        $\pi_{s}^*$ on $P_t$    & $-1076.92 \pm 1597$           & $-281.33 \pm 70$          & $-271.66 \pm 2$           & $486.1 \pm 23.33$\\ 
        \hline
        $\pi_t$ on $P_t$        & $\mathbf{-904.02 \pm 1624}$   & $-2.50 \pm 3$             & $-7.24 \pm 8$             & $\mathbf{500.00 \pm 0}$\\
        \hline
        $\pi_t^-$ on $P_t$      & $\mathbf{-904.02 \pm 1624}$   & $\mathbf{-2.46 \pm 3}$    & $\mathbf{-7.16 \pm 8}$    & $\mathbf{500.00 \pm 0}$\\
        \hline
        $\pi_t^+$ on $P_t$      & $\mathbf{-904.02 \pm 1624}$   & $-2.48 \pm 3.14$          & $-7.17 \pm 8$             & $\mathbf{500.00 \pm 0}$\\
        \hline
        \end{tabular}
        \caption{Policy transformation with known $F_A, F_B$.}
        \label{tbl:exp_rl_known}
    \end{subfigure}
    \begin{subfigure}{\textwidth}
        \centering
        \begin{tabular}{lllll} 
        \hline
                                & Temperature                   & Spring                    & Pendulum              & Cartpole \\ 
        \hline
        $\pi_s$ on $P_t$        & $-51123.05 \pm 17246$         & $-295.39 \pm 70$          & $-271.66 \pm 2$       & $56.8 \pm 15.16$\\ 
        \hline
        $\pi_{s}^*$ on $P_t$    & $-1076.92 \pm 1597$           & $-281.33 \pm 70$          & $-271.66 \pm 2$       & $486.1 \pm 23.33$\\ 
        \hline
        $\pi_t$ on $P_t$        & $\mathbf{-904.02 \pm 1624}$   & $-171.15 \pm 175$         & $-19.84 \pm 7$        & $156.5 \pm 88.67$\\
        \hline
        $\pi_t^-$ on $P_t$      & $-904.33 \pm 1624$            & $\mathbf{-110.60 \pm 169}$& $-12.79 \pm 8$        & $\mathbf{500.00 \pm 0}$\\
        \hline
        $\pi_t^+$ on $P_t$      & $\mathbf{-904.02 \pm 1624}$   & $-134.11 \pm 168.62$      &$\mathbf{-10.96 \pm 8}$ & $493.8 \pm 18.6$\\
        \hline
        \end{tabular}
        \caption{Policy transformation with unknown $F_A, F_B$. They are approximated using a buffer of measurements for 5 episodes of $\pi_s$ on $P_t$.}
        \label{tbl:exp_rl_unknown}
    \end{subfigure}
    \caption{Mean episodic rewards using RL.}
    \label{tbl:exp_tables_rl}
\end{table*}

%% file: sections/conclusions.tex
\section{Discussion and Conclusion}

The results demonstrate several key points. Our approach gets a jump-start improvement in peformance after a parametric fault. Secondly, during RL, if not already converged, rewards are faster to reach convergence threshold (figure \ref{fig:learning}). Thirdly, when knowledge of transformation and dynamics is known, the source policy's transformation gives results similar to LQR and MPC being trained on $P_t$ (tables \ref{tbl:exp_tables_classical}, \ref{tbl:exp_rl_known}). However, unlike MPC, an optimization problem does not need to be solved recurrently when applying control, saving computational complexity. Finally, when knowledge of transformation and dynamics is not known, an approximate transformation using measured samples and fine-tuning using RL gives similar, albeit marginally lesser, results (table \ref{tbl:exp_rl_unknown}).

Therefore, our approach may lend itself as an initialization strategy for data-driven controllers to mitigate sample inefficiency. After the adaptation step, the controller can proceed with further reinforcement learning to fine-tune parameters.

We looked at transforming control policies by reasoning about task dynamics as a means of adaptive control, instead of the statistical properties of parameters as in machine learning. Our main contribution was the transformation of a nominal control policy that leverages system identification. It is applicable to a host of control algorithms, and tasks where the objective function is agnostic to actions. The transformation is such that a source policy would transfer positively on the target process with a higher sample efficiency than reinforcement learning.

There are several interesting venues of future research. First, using the error in policy transformation, which may use pseudo-inverses, as a measure and guarantee of the quality of transfer. Secondly, extending the transformation to a broader class of MDPs with non-linear disturbances. Thirdly, using parameter estimation, by relying on fault identification, instead of system identification  to further reduce the data samples to adapt to a new task.